# Orbital Angular Momentum Microlaser with Tunable Degree of Chirality and Large Quantum Numbers

Zhen Qiao[1], Song Zhu[1], Yu-Cheng Chen[1, *]

[1]School of Electrical and Electronic Engineering, Nanyang Technological University, 50 Nanyang Ave., Singapore 639798, Singapore

* Correspondence Email: *yucchen@ntu.edu.sg*

**Abstract**

Chiral microlasers with orbital angular momenta (OAM) are promising platforms for developing integrated high-dimensional photonic devices. However, it's still challenging to arbitrarily manipulate the degree of chirality (DOC) and quantum numbers of OAM. Herein, we proposed and demonstrated OAM generations with tunable DOC and large quantum numbers from microlasers. Through the design of an optical microcavity with uneven potential distribution of photons, the dissymmetry factor of OAM laser can be continuously tuned from -1 to +1 by governing the pump position. High-order vortices were also generated, in which the largest quantum number reached up to 352. We further demonstrated multi-vortex laser generation in spatial and temporal domains. This study provides a novel mechanism for manipulating the chirality of OAM based on microlasers, with a great potential in high-dimensional information processing.

**Introduction**

Optical vortices are helical modes of light which are ubiquitous in modern science for optical communications and quantum optics. Such vortex beams are notable for their unique capability of carrying orbital angular momentum (OAM) of $l\hbar$ per photon ($l$: quantum number or topological charge, and $\hbar$ is Planck's constant)[1]. Throughout the years, attempts have been made to generate OAM laser directly from micro-resonators due to their huge potential for developing miniaturized integrated laser devices. Namely, chiral-structured resonators[2, 3], non-Hermitian systems[4-6], quantum Hall effect[7] and spin-orbit coupling[8] have been employed to construct vortex microlasers. Due to the infinite quantum numbers of OAM, optical vortices provide higher-dimensional quantum states serving as powerful sources for information encoding[9-14], ultrasensitive detection[15], and investigation of fundamental quantum mechanics[16-18].

Orbital angular momentum laser possesses multiple optical degree of freedoms for manipulation in both classical and quantum optical regimes[19], including polarization[20], wavelength[21, 22], pulse duration[23, 24], quantum number[25], and chirality[8]. Take for example, manipulating the chirality of OAM laser will efficiently scale the quantum states[8]. In principle, vortex modes with opposite chirality can be simultaneously excited due to symmetry restriction of micro-resonators, resulting in the emission of superposition OAM states $C_+|+l\rangle+C_-|-l\rangle$. The difference between the normalized coefficients $|C_+|^2$ and $|C_-|^2$ is the dissymmetry factor which represents the degree of chirality (DOC) of OAM laser[26]. Arbitrarily tuning the DOC may further expand the information processing capacity. However, the ability to fully and flexibly manipulate the DOC of OAM remains elusive under microscale. On the other hand, most micro-resonators only allow small quantum numbers with less scalability. It's still an ongoing challenge to create OAM with arbitrary DOC and arbitrary large quantum numbers at point of microlaser sources.

In this study, we demonstrated the generation of OAM with tunable DOC and large quantum numbers through microlasers. As illustrated in **Figure 1**, OAM laser was generated through a Fabry–Pérot (FP) microcavity comprised of a pair of tilted mirrors. A microring structure was fabricated on the bottom mirror before introducing optofluidic dyes as the laser gain medium. The flow directions of photons along the microring can be governed by controlling the excitation region of the microcavity. As such, we realized the arbitrary manipulation of DOC of OAM laser by tuning the pump positions. Furthermore, vortex modes with a wide range of quantum numbers (5~352) were obtained by using microrings with various diameters. Finally, we demonstrated the simultaneous emission of multiple vortex modes and dynamic tuning of chirality based on multiple microrings and pump beams. This work gives a simple method to manipulate the chirality and quantum numbers of OAM modes at the source, which significantly improved the tunability of chiral microlasers.

**Results**

*Concept and Principle*

**Figure 1** illustrates the concept and principle for generating OAM with tunable degree of chirality (DOC) from a microlaser. A Fabry–Pérot (FP) microcavity was utilized to construct a microlaser, which is composed of two distributed Bragg reflector (DBR) mirrors. Optofluidic fluorescent dye was encapsulated in the microcavity to perform as the laser gain. As shown in **Fig. 1a**, a microring structure was fabricated on the bottom mirror (**Supplementary Fig. S1**).[27] Such a polystyrene (PS) microring possess a higher refractive index compared with the surrounding environment, which can confine photons in the ring region and supports the oscillations of OAM modes. More importantly, a small tilting angle was formed between the DBR mirrors. The purpose was to break the symmetry restriction of a microcavity. Now assuming that two vortex modes with opposite chirality oscillate simultaneously in the microcavity, the emission laser will carry the superposition OAM states of

$$|\psi\rangle = C_+|+l\rangle + C_-|-l\rangle \tag{1}$$

where $\pm l$ denotes the quantum number (or topological charge), and the $C_+$ and $C_-$ are normalized coefficients ($|C_+|^2 + |C_-|^2 \equiv 1$). Here we define the dissymmetry factor of OAM laser to quantify the DOC, which is written as the following equation:

$$g_{OAM} = |C_+|^2 - |C_-|^2 \tag{2}$$

Due to the fact that two DBR mirrors could not be perfectly parallel, the dissymmetry factor $g_{OAM}$ can be arbitrarily tuned in the range of [-1, +1] by controlling the pump position on a microring (**Fig. 1a**). **Figure 1b** shows the morphology of a fabricated PS microring, where the AFM image indicates that the height of the microring was at nano-meter scale.

We further performed the theoretical analysis and simulations on the generation of OAM laser with tunable DOC, as shown in **Figs. 1c-1g**. The side- and top-views of a microcavity is illustrated in **Figs. 1c-1d**, in which a microring and a small tilting angle between the mirrors were considered. Such a microcavity possesses an effective potential for confined photons [28] (**Supplementary Discussion I**):

$$V(\theta) = -\frac{c^2 D \alpha}{n_0^2 L_0} \sin^2\left(\frac{\theta}{2}\right) \qquad (3)$$

where $c$ is the velocity of light; $D$ is the diameter of a microring; $n_0$ is the refractive index of the gain medium in the microcavity; $L_0$ is the cavity length; $\alpha$ is the tilting angle between the mirrors; and $\theta$ is the azimuth angle with respect to the tilting axis of the microcavity (**Fig. 1d**). According to Eq. (3), the potential is depicted in **Fig. 1e**, which exhibits a "hill-shaped" distribution as a function of azimuth angle $\theta$. Such a potential distribution allows one to manipulate the DOC of OAM laser by controlling the pump position on a microring. As illustrated in **Figs. 1d and 1e**, when pump at a specific azimuth angle $-\theta_1$ (or $+\theta_2$) on a microring, the excited photons will flow along the clockwise (or counter-clockwise) direction due to reflections by the unparallel mirrors, which results in the oscillation of a vortex mode $|-l\rangle$ (or $|+l\rangle$). As such, the dissymmetry factor can be arbitrarily tuned from $-1$ to $+1$ by adjusting the pump position in the range of $[-\theta_1, +\theta_2]$. If the pump position is deviated from the range of $[-\theta_1, +\theta_2]$, the oscillated mode will not form a closed ring (**Supplementary Fig. S2**).

**Figures 1f and 1g** show the simulated electric-field distributions and spatial phases of the oscillated OAM modes under different pump positions marked in **Fig. 1d**. When pump at the azimuth angle of $-\theta_1=-47°$, a clockwise-vortex mode $|-3\rangle$ is generated. In contrary, a counter-clockwise-vortex mode $|+3\rangle$ is generated under the pump position of $+\theta_2=50°$. If the pump spot is located right on the tilting axis of the microcavity, the two vortex modes $|-3\rangle$ and $|+3\rangle$ both oscillate. In this case, nodal points can be observed in the laser pattern around the ring (middle part in **Fig. 1f**), which is originated from the interference between the two vortex modes. Note that, a vortex mode possesses a non-uniform intensity distribution and unequally phase distribution around the center due to the unparallel mirrors. It's worth mentioning that, in the simulations and the following experiments, silver nanocubes were applied to spread on a microring. The nanoparticles introduce more loss in a microcavity, which contribute to enhance the DOC of OAM laser under a fixed pump position. In general, $\theta_1$ and $\theta_2$ are slightly different under a non-uniform loss distribution. Details on the simulation can be referred to **Supplementary Fig. S3**.

*OAM laser generation with tunable DOC*

In experiment, we first demonstrated the generation of vortex modes with opposite chirality, i.e., $|-l>$ and $|+l>$. A microring with the diameter of ~120 μm was fabricated on a DBR mirror (**Fig. 2a**). In addition, silver nanocubes with the side length of 100 nm were coated on the microring, which can be observed in the dark-field image (right in **Fig. 2a**). Thereafter, a microcavity was built by covering the top DBR mirror and encapsulating fluorescein sodium salt (FITC) dye as the gain material. The tilting angle between the DBR mirrors was measured with optical interference method. After illuminating the microcavity with monochromatic light, interference fringes can be clearly observed in the far-field plane (**Fig. 2b**). By measuring the spatial period of the fringes, the tilting angle between the mirrors was calculated as 0.02°. The "opening" direction of the mirrors (along *X*-axis marked in **Fig. 2b**) was judged by pumping at a blank region of the bottom mirror, for which the oscillated laser also tends to flow along *X*-axis (**Fig. 2c**).

By applying above mentioned microcavity, vortex modes with opposite chirality were generated under different pump positions. **Figures 2d and 2e** show the laser patterns of vortex modes $|-91>$ and $|+91>$, where the white dash circles denote the pump regions. Narrow-linewidth lasing peaks in the optical spectra show typical lasing actions of microlasers (**Figs. 2f and 2g**). The spectral integrated intensity as a function of the pump energy density present clear threshold behaviors, further indicating the lasing operations. The quantum number (or topological charge) was measured by converting a vortex mode into a long-stripe mode using a cylindrical lens, where the number of nodal lines in the converted laser pattern is equal to the absolute value of quantum number[29] (**Supplementary Fig. S4**).

We further demonstrated that the DOC of OAM laser can be arbitrarily tuned by controlling the pump positions. When continuously adjusted the pump position in the range of [-15°, 20°] on the above-mentioned microring, the dissymmetry factor gradually evolved from -1 to +1 (**Fig. 3**). **Figure 3b** shows the OAM laser patterns under different pump positions, which present the process of chirality inversion. The dissymmetry factor was measured by converting the OAM laser with a cylindrical lens. For the superposition OAM states of $|+l\rangle$ and $|-l\rangle$, the converted laser pattern presents an "X-shaped" distribution, which is composed of two long-striped modes with different longitudinal directions (**Fig. 3c**). The intensity difference between the two long-striped modes were measured to obtain the dissymmetry factor (**Supplementary Fig. S4**). We also studied the photon energy flows by using two coupled microrings, where the energy-flow direction of OAM laser can be clearly reflected by the electric fields coupled in the adjacent microring (**Supplementary Fig. S5**).

The proposed microcavity structure also possesses the ability to scale quantum numbers in a wide range. In principle, an arbitrary quantum number can be obtained by tuning the diameter of a microring. For experimental demonstrations, we fabricated microrings with various diameters (27 μm~350 μm) and integrated them into FP microcavities. Subsequently, high-purity vortex modes with the quantum numbers of 5 ~ 352 were obtained (**Fig. 4**). It's worth mentioning that, 352 is the largest quantum number of a vortex mode generated from laser up to now. Note that, the polarization state of a vortex mode is linearly polarized (**Supplementary Fig. S6**).

*Multi-vortex laser generation*

In general, OAM modes with various quantum numbers are required for practical applications, especially for information processing based on space-division multiplexing and temporal encoding. In order to show the potential of the chiral microlasers for developing integrated devices, here we demonstrated simultaneous generation of multiple vortex modes and dynamic tuning of chirality. **Figure 5a** illustrates the schematic of simultaneous generation of multiple vortex modes on chip. Two microrings with different diameters were fabricated adjacently on a DBR mirror, which were excited by two synchronous pump beams (**Fig. 5b**). Subsequently, vortex modes with different quantum numbers were simultaneously generated from laser. The laser patterns of the two vortex modes are shown in **Fig. 5c**. The quantum numbers were measured to be $l=-20$ and $l=28$, respectively.

As a proof-of-concept, we further performed the dynamic tuning of chirality based on a fixed microring. As shown in **Figs. 5d and 5e**, the chirality of vortex laser was dynamically switched by exciting two different positions on a microring with two asynchronous pulsed pump beams. A single pump pulse resulted in the emission of a vortex mode with a specific chirality. As such, the chirality of the output vortex laser was dynamically switched (**Fig. 5f**). Technically, the modulation speed can be significantly improved by applying pump pulses with higher repetition rates. Ultrafast modulation can also be achieved if picosecond or femtosecond pulses are used[30]. The experimental details and measurements of the quantum numbers are shown in **Supplementary Figs. S7.**

**Discussions and Conclusions**

Here we would like to discuss the effect of silver nanocubes applied in this study. Silver nanocubes spread on a microring will provide more loss before the oscillation is built due to plasmonic absorption and scattering. Simulations and experimental results both show that the loss effect will enhance the DOC of OAM laser (**Supplementary Fig. S8**). When pump on a position deviated from the tilting axis of a microcavity, most excited photons will propagate along one direction (e.g., clockwise direction) due to the uneven potential of a microcavity, which contribute to the oscillation of clockwise-vortex mode. Meanwhile, a small part of the excited photons can still pass through the potential barrier if little loss is introduced, and thereafter participate in the oscillation of counter-clockwise-vortex mode. As such, two opposite vortex modes will be excited simultaneously, resulting in the emission of OAM laser with a relatively small dissymmetry factor. However, when introduce more loss, that small part of excited photons which propagate along counter-clockwise direction will be suppressed. Thereby, the DOC can be efficiently enhanced. In addition to silver nanocubes, we can also introduce more loss in other forms. For example, we fabricated a microring doped with Rhodamine 6G (Rh6G) dye which provide more absorption loss for FITC-based laser. As expected, the emitted OAM laser also possesses a high dissymmetry factor under a specific pump position (**Supplementary Fig. S10**). The chirality manipulation can also be understood based on a non-Hermitian two-level quantum system, in which the loss and gain co-determine the dissymmetry factor of OAM laser[4].

The manipulation of DOC has the potential in high-capacity information processing. Similar to fractional vortex modes which can infinitely divide the topological charges between two neighborhood integers[31, 32], the superposition vortex modes with opposite chirality also possess infinite values of dissymmetry factors. Thus, a higher-capacity optical communication may be achieved by applying DOC as the encoding dimension. The developed microlasers also have the advantages of flexibility and tunability. Different from WGM-based microring cavities, an FP microcavity has the intrinsic ability to emit laser with a high directionality, which do not need high-precision output coupling structures. Complex photonic structures are either not required to manipulate the chirality and quantum number. Based on the microcavities proposed in this work, the chirality is totally optically controlled, and the quantum number can be tuned by adjusting the microring diameter. Moreover, the laser gain can be extended to different materials, such as other organic materials, Perovskite, quantum dots, etc. As such, the laser wavelength can also be adjusted.

In summary, a new type of chiral microlaser was developed for which the DOC and quantum number of OAM laser can be arbitrarily tuned. The proposed microcavity structure allows one to fully control the energy-flow direction of the confined photons. As such, the arbitrary tuning of DOC was achieved. In addition, the quantum numbers were scaled in a range of two orders of magnitude. Finally, on-chip manipulation of multiple vortex modes in spatial and temporal dimensions were also performed. By taking the advantage of high tunability, the proposed microlaser scheme possesses a great potential for developing minimized high-dimensional OAM laser systems.

**Materials and Methods**

*Materials*

FITC, PS, N, N′-Dimethylformamide (DMF), and toluene were purchased from Sigma-Aldrich. Silver nanocubes were purchased from Nano Composix. FITC solution was prepared by dissolving FITC dye in DI water.

*Fabrication of PS microrings*

The schematic of the fabrication process is shown in **Supplementary Fig. S1**. First, a solution of PS in toluene with the concentration of 30~60 mg/mL was prepared. A PS film was then spin-coated on a DBR mirror with the rotating speed of 3000 rpm. Thereafter, DMF droplets were dispersed onto the mirror to etch ring-shaped structures at 80°C, forming PS microrings with various diameters. For the fabrication of silver-nanocube layer, a dispersion of silver nanocubes (100 nm-side length) in ethanol was coated on the mirror. Subsequently, a silver-nanocube layer was formed on the PS film after the evaporation of the ethanol.

*Fabrication of a microcavity*

A FP microcavity was fabricated by using a pair of DBR mirrors, in which the bottom mirror was integrated with PS microrings. A 4 mM FITC solution was employed as the gain material. Microbeads with 20 μm-diameter were used as spacers to fix the cavity length. The small tilting angle between the two mirrors were formed by clamping two ends of the cavity with different forces.

*Optical setup*

The optical system is illustrated in **Supplementary Fig. S4**. The pump source was a pulsed ns-laser (EKSPLA NT230) integrated with an optical parametric oscillator (repetition rate: 50 Hz; pulse duration: 5 ns). An upright microscopic system (Nikon Ni2) with 10 × 0.3 NA objective was used to focus the pump light into a FP microcavity, and was also used to collect the emitted laser. The pump wavelength was 480 nm. A charge-coupled device (CCD) camera (Newton 970 EMCCD) was used to record the laser patterns. An imaging spectrometer (Newton KYMERA-328I-D2) was used to record the lasing spectra. For the measurement of quantum numbers and dissymmetry factors, a cylindrical lens was inserted at a proper position between the objective and the CCD to convert the OAM modes into long-stripe modes.

*Simulations*

The simulations on the generation of OAM modes in **Fig. 1** were performed by using finite element method using Comsol Multiphysics software. The Eigenfrequency Study was applied in the Electromagnet Waves, Frequency Domain interface within the Wave Optics modules, in which 3D model was employed. The pump region is simulated by setting the imaginary part of refractive index of a region as a negative number (**Supplementary Fig. S3**).

**Acknowledgements**

All the authors would like to thank A*STAR for support. This research is supported by A*STAR under its AME IRG Grant (project no. A20E5c0085). M. K. is also thankful for the support from ACRF Tier 2 (Grant T2EP50120-0003) and the lab support from the

Nanyang NanoFabrication Centre.

**Author Contributions**

Z. Q. and Y. -C. C. conceived the research and designed the experiments; Z. Q. performed the experiments and the theoretical calculations; Y. L. and M. K. conducted the AFM measurement; S. Z., C. G., Z. Y. and X. G. participated the discussions on the project. Z. Q. and Y. -C. C analyzed the data and co-wrote the paper.

**Additional information**

The authors declare no competing financial interests.

Supplementary materials are available upon request.

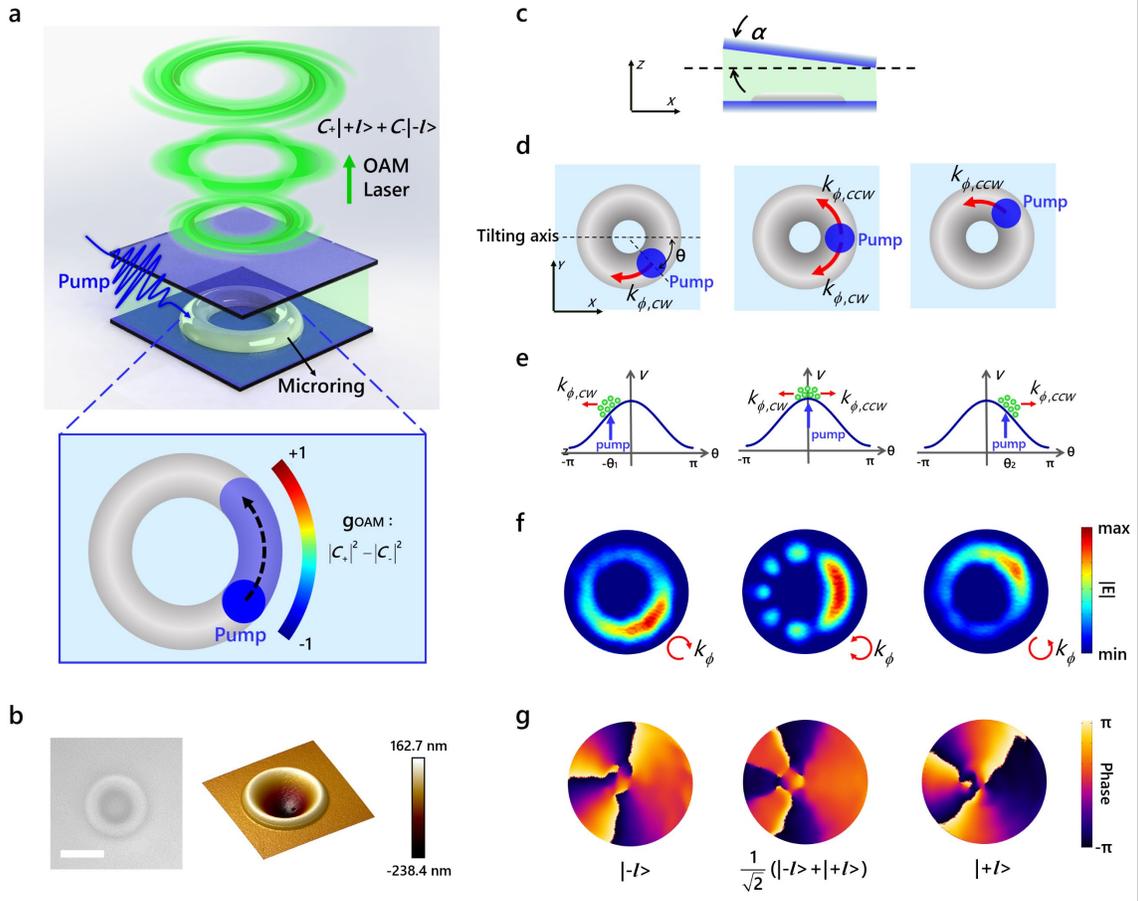

**Figure 1. OAM laser generation with arbitrarily tunable degree of chirality. (a)** Illustration of a chiral microlaser, where a microring structure and gain material were integrated in a Fabry–Pérot (FP) microcavity. OAM laser with a specific dissymmetry factor ($g_{OAM}$) will emit when pump on a specific position of the microring. DBR, distributed Bragg reflector. **(b)** Bright-field image (Left) and atomic force microscopic (AFM) image (Right) of a fabricated microring. Scale bar: 25 μm. **(c)-(g)** Principle and simulation of OAM laser generation with tunable DOC. **(c)** Schematic illustration of a FP microcavity in *X-Z* plane, where a small tilting angle was formed between the DBR mirrors. **(d)** Schematic illustration of the FP microcavity in *X-Y* plane. Blue spots denote the pump regions. $k_{\phi,cw}$ and $k_{\phi,ccw}$ denote the wavevectors of photons propagating along the clockwise and counter-clockwise directions in *X-Y* plane. **(e)** Effective potential *V* of a microcavity. The energy flow of excited photons can be tuned by controlling the pump position on the microring. **(f)** Electric-field-strength distributions of OAM modes with different DOC under the pump positions in (d). **(g)** Spatial phase distributions of the OAM modes in (f).

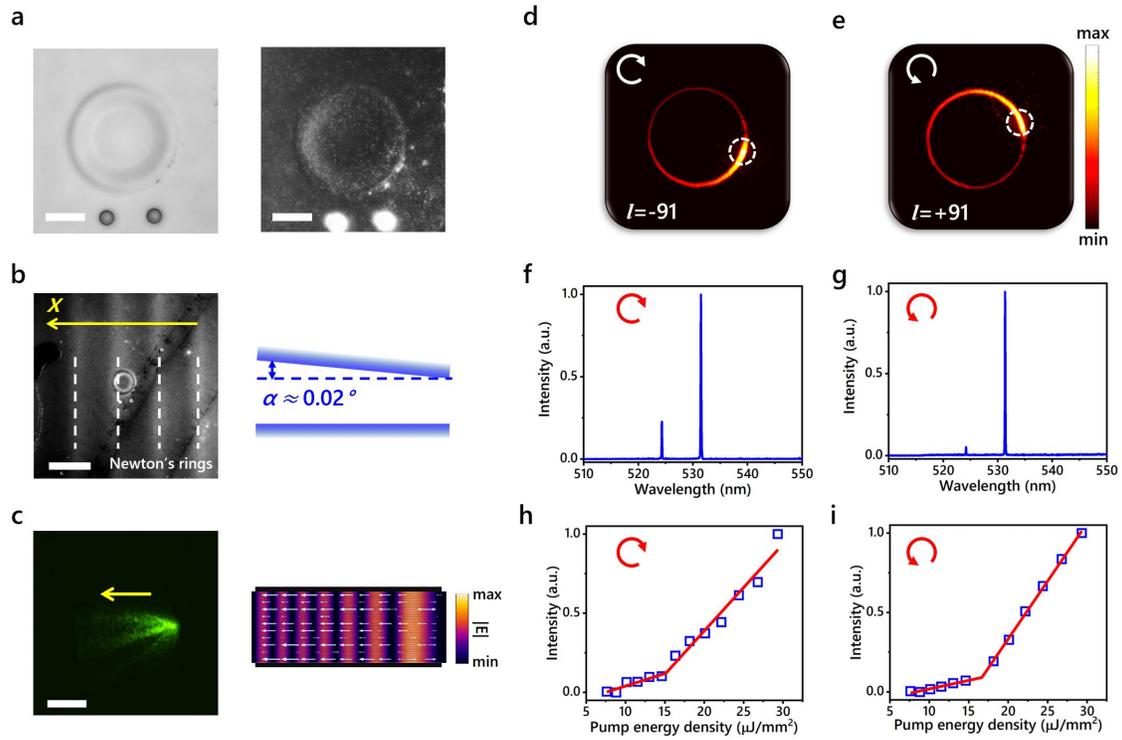

**Figure 2. Generation of vortex modes with opposite chirality.** **(a)** Bright-field image (Left) and dark-field image (Right) of a microring. **(b)** Left: Bright-field image of the microcavity illuminated by monochromatic light, where Newton's rings can be observed due to the unparallel mirrors. Right: Schematic illustration of the microcavity in side view. The tilting angle was calculated as 0.02° according to the Newton's rings. **(c)** Left: Laser pattern by pumping on a blank region of the bottom mirror, which shows the energy flow along $X$-axis. Right: Simulated electric-field-strength distribution in the microcavity, where the white arrows denote Poynting vectors along $X$-axis. **(d) (e)** Far-field laser patterns of vortex modes $|-91\rangle$ (d) and $|+91\rangle$ (e) under different pump positions (white dashed circles). **(f) (g)** Laser spectra of the vortex modes in (d) and (e). **(h) (i)** Spectral integrated intensity as a function of pump energy density of the vortex modes in (d) and (e). Scale bars in (a) (d) and (e): 50 μm. Scale bars in (b) (c): 400 μm.

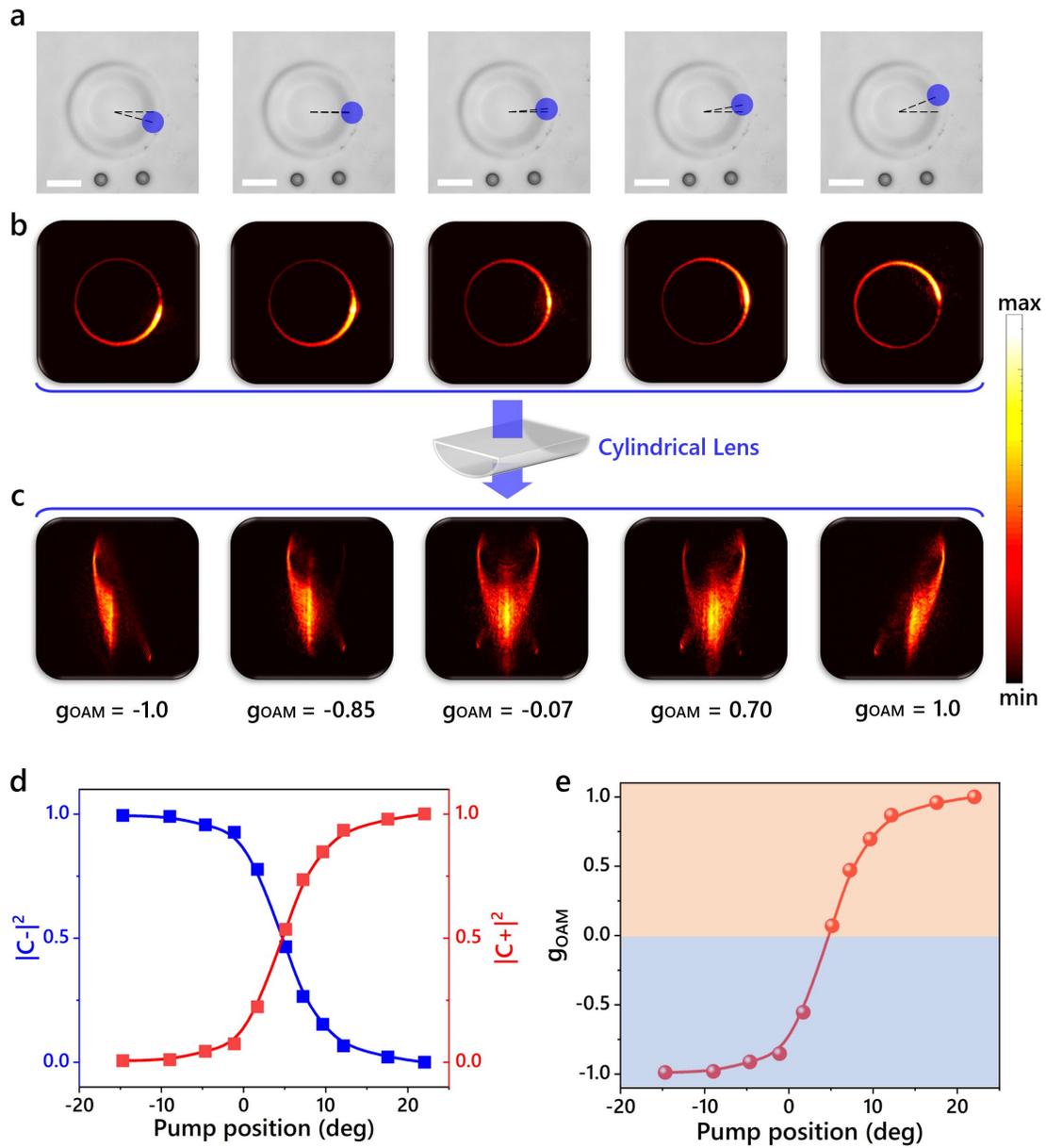

**Figure 3. Tunable DOC of OAM laser. (a)** Bright-field images of the microring. Blue spots denote the pump regions. Scale bar: 50 μm. **(b)** Far-field laser patterns of the output OAM modes with various DOC under the pump positions in (a). **(c)** Laser patterns after conversion using a cylindrical lens. **(d)** Measured purity of $|+l\rangle$ and $|-l\rangle$ as a function of pump position on the microring. **(e)** Measured dissymmetry factor $g_{OAM}$ as a function of pump position on the microring.

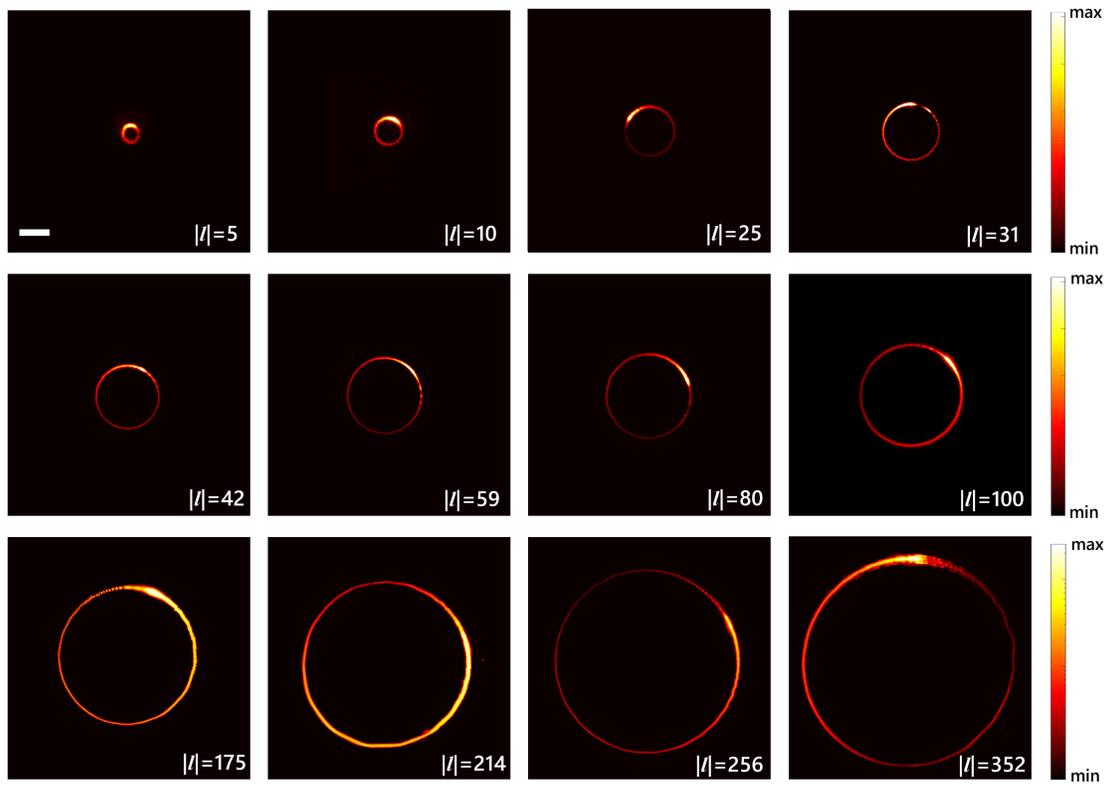

**Figure 4. Generation of vortex modes with various quantum numbers.** Far-field laser patterns of vortex modes with the quantum numbers of $|l|$=5, 10, 25, 31, 42, 59, 80, 100, 175, 214, 256, and 352. Scale bar: 50 μm

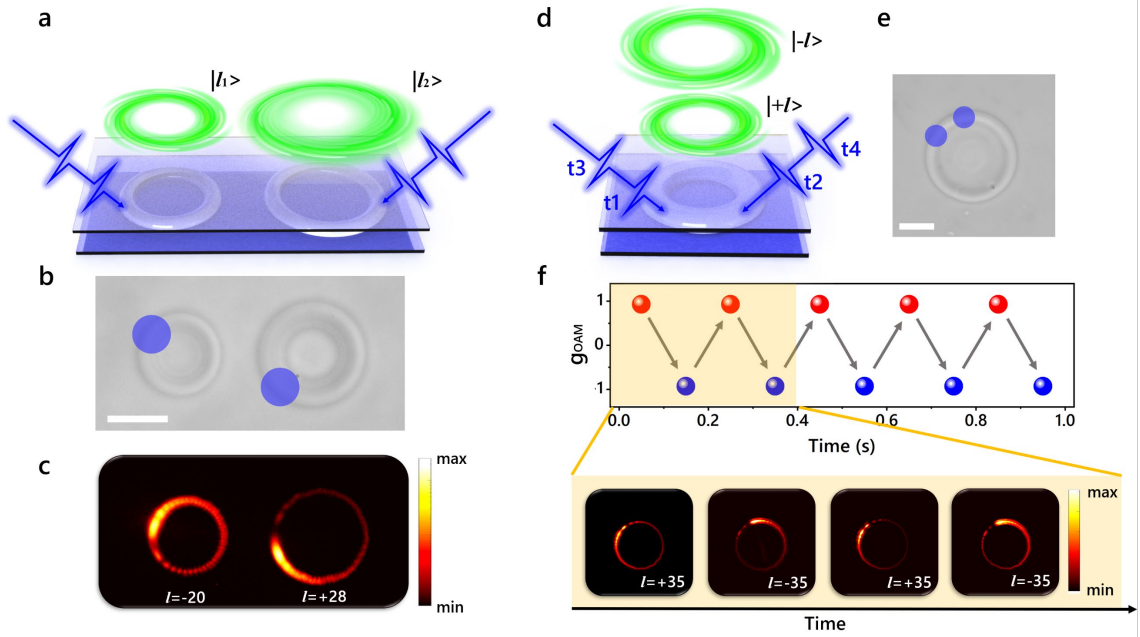

**Figure 5. Multi-vortex laser generation**. **(a)-(c)** Simultaneous generation of multiple vortex modes. **(a)** Schematic illustration of a microlaser integrating two microrings with different diameters. By exciting the two microrings with two synchronous pump beams, vortex modes with different quantum numbers will simultaneously emit from the microlaser. **(b)** Bright-field image of the microrings. Blue spots denote the pump regions. **(c)** Far-field laser patterns of the simultaneously emitted vortex modes. **(d)-(f)** Dynamically tunable chirality of vortex laser. **(d)** Schematic illustration of a microlaser. By pumping on two different positions of a microring with two asynchronous pump beams, the microlaser emits vortex modes with dynamically switching chirality. **(e)** Bright-field image of a microring. Blue spots denote the pump regions. **(f)** Top row: dissymmetry factor $g_{OAM}$ as a function of time. Below row: Far-field laser patterns of the vortex modes at different time. Scale bars in (b) and (e): 50 μm.